\documentclass[12pt,preprint]{aastex}
\usepackage{color}

\shorttitle{superflare ICME}
\shortauthors{Takahashi et al.}

\begin{document}
\title{Sheath-Accumulating Propagation of Interplanetary Coronal Mass Ejection}

\author{
Takuya Takahashi\altaffilmark{1},
Kazunari Shibata\altaffilmark{1}}
\email{takahasi@kusastro.kyoto-u.ac.jp}

\altaffiltext{1}{
Kwasan and Hida Observatories, Kyoto University,
Yamashina, Kyoto 607-8471, Japan.}

\begin{abstract}
Fast interplanetary coronal mass ejections (interplanetary CMEs, or ICMEs) are the drivers of strongest space weather storms such as solar energetic particle events and geomagnetic storms. The connection between space weather impacting solar wind disturbances associated with fast ICMEs at Earth and the characteristics of causative energetic CMEs observed near the Sun is a key question in the study of space weather storms as well as in the development of practical space weather prediction. Such shock-driving fast ICMEs usually expand at supersonic speed during the propagation, resulting in the continuous accumulation of shocked sheath plasma ahead. In this paper, we propose the ``sheath-accumulating propagation'' (SAP) model that describe the coevolution of the interplanetary sheath and decelerating ICME ejecta by taking into account the process of upstream solar wind plasma accumulation within the sheath region. Based on the SAP model, we discussed (1) ICME deceleration characteristics, (2) the fundamental condition for fast ICME at Earth, (3) thickness of interplanetary sheath, (4) arrival time prediction and (5) the super-intense geomagnetic storms associated with huge solar flares. We quantitatively show that not only speed but also mass of the CME are crucial in discussing the above five points. The similarities and differences among the SAP model, the drag-based model and the`snow-plough' model proposed by \citet{tappin2006} are also discussed.
\end{abstract}


\section{INTRODUCTION}
Coronal mass ejections (CMEs) are the largest plasma explosions in the solar system, where a vast amount (typically, $10^{13}-10^{16}$~g) of the solar coronal plasma is ejected out into the interplanetary space with speeds up to 3000 km~s$^{-1}$ \citep{illing1986,aarnio2012,porfireva2012,webb2012}. 

CMEs propagating in the interplanetary space are called interplanetary CMEs or ICMEs. When the magnetic field of ICME observed in-situ shows continuous rotation, it is called magnetic cloud \citep{klein1982}. Fast ICMEs drive shock waves. The draped plasma compressed and accelerated by the leading shock is called interplanetary sheath.
The sudden jumps in energetic proton flux measured in-situ associated with the passage of interplanetary shocks are called energetic storm particle (ESP) events \citep{bryant1962}. On the other hand, when the magnetic cloud or interplanetary sheath that arrive at Earth posses southward magnetic field, they drive intense geomagnetic storms \citep{dungey1961,klein1982,tsurutani1988,zhang2004}.

One super-intense ESP event was recently recorded on 2012 July 23 by $STEREO-A$ spacecraft located at 0.96 AU from the Sun and at 121$^{\circ}$ ahead of Earth at the time. The estimated peak E$~>10~$MeV proton flux measured in-situ by $STEREO-A$ was about $6.5\times10^4$ pfu when the interplanetary shock passed the spacecraft \citep{russell2013,gopalswamy2014}. \citet{liu2014} carried out comprehensive study on the ICME characteristics with stereoscopic observations and in-situ measurements. The post-shock peak speed measured in-situ at $STEREO-A$ is about 2250 km s$^{-1}$. The recorded peak magnetic field strength is 109 nT, being among the largest interplanetary field strength on record near 1 AU. \citet{liu2014} concluded that we would have had an extreme geomagnetic storm with minimum DST index \footnote{http://wdc.kugi.kyoto-u.ac.jp/dstdir/} of $Dst\sim-1150$ or $-600$ nT had the ICME arrived at Earth, based on two different empirical formula. The causative CME resulted from the merger of two successive CMEs with the peak speed of about 3050 km s$^{-1}$ near the Sun. Also, a fast CME preceded the July 23 event by 4 days from the same solar active region, and the solar wind trailing the preceding CME had its density as low as 1 cm$^{-3}$. They discussed that both the ``preconditioned'' low-density upstream solar wind and the CME merger have played a crucial role in producing very fast ICME near 1AU with extremely strong magnetic field.

Taking into account the fact that ICMEs faster than 1000 km s$^{-1}$ at 1 AU are extremely rare \citep{guo2010}, the ICME of 2012 July 23 event had a truly outstanding propagation characteristics. What is the fundamental condition for such an extremely space weather impacting ICME at 1 AU remains an issue of crucial importance for space weather research.

Recently, superflares (flares that are 10-1000 times more energetic than the largest ever observed solar flares) on solar-type stars were discovered with Kepler data \citep{maehara2012}. The possible impacts of superflares on space weather and terrestrial environment are also discussed vigorously \citep{miyake2012,schrijver2012,shibata2013,hayakawa2015,tsurutani2014,airapetian2016,takahashi2016}. 

A landmark for realistic space weather prediction is to connect the coronagraph observation of a CME near the Sun with an expected ICME characteristics at 1 AU, as well as its arrival time. As for the prediction of the arrival time of ICME at 1 AU, various methods have been proposed so far \citep{dryer1984,smith1990,cargill1996,feng2006,gopalswamy2000,gopalswamy2005}.
One of the most typical models is the empirical CME arrival (ECA) model proposed by \citet{gopalswamy2000}. The ECA model assumes ``effective'' constant deceleration (or acceleration) of CME in the interplanetary space. \citet{gopalswamy2001} introduced deceleration cessation distance of 0.76AU after which ICME propagate with constant speed in order to improve predictability of ECA model. The empirical shock arrival (ESA) model developed by \citet{gopalswamy2005} based on the combination of ECA model prediction and piston-driven shock propagation predicts 1 AU arrival time of leading shock front. On the other hand, the drag-based model predicts 1 AU arrival time of CME ejecta taking into account the deceleration of CME by aerodynamic (or viscous) drag \citep{cargill1996,cargill2004,vrsnak2001,vrsnak2010,vrsnak2013}. 
The `snow-plough' model proposed by \citet{tappin2006} discussed the deceleration of CME based on the conservation of momentum as the CME sweeps up the slower solar wind ahead of it.

The actual CME propagation in highly structured solar wind plasma involves various dynamical processes, such as deflection and rotation of CME flux rope, magnetic reconnection between CME and ambient solar wind, deformation of flux rope, etc. They are studied vigorously with the use of global MHD simulation in 3D \citep{manchester2004,lugaz2011,shiota2016}.

\citet{liu2013} reported the first detailed examination of Sun-to-Earth propagation characteristics of thee fast CMEs and associated shock front with the combination of wide angle heliospheric imaging observation by $STEREO$, interplanetary Type II radio bursts, and in-situ observation of solar wind parameters in multiple points. They reported that CME Sun-to-Earth propagation is approximately formulated into three phases, that are (1) an impulsive acceleration near the Sun (upto $\sim{}0.1$ AU), (2) rapid deceleration up to the distance of $\sim{}0.2-0.4$ AU and (3) nearly constant speed propagation or gradual deceleration afterwards. 

Shock driving fast ICMEs are accompanied with continuously accumulating sheath plasma ahead. 
We need a model that describes the coevolution of interplanetary sheath and shock-driving ICME propagation on a single theoretical basis. Such a model would be helpful for the understanding of space weather impacting ICME-related disturbances at 1 AU such as speed, magnetic field strength and size of both interplanetary sheath and magnetic cloud as well as their arrival times. For this purpose, we construct a new model that connects ICME deceleration and interplanetary shock propagation by taking into account the process of upstream solar wind plasma accumulation within the sheath region. We call the model ``sheath-accumulating propagation'' (SAP) model of ICMEs. We note that the effect of the Lorentz force and gravity are neglected in the SAP model, which could also be effective especially in the vicinity of the Sun \citep{chen2010}.

In section 2, we introduce the SAP model and investigate the ICME propagation characteristics of the SAP model. In section 3, we discuss the thickness of interplanetary sheath ahead of ICME. In section 4, we compare the SAP model with the drag-based model. In section 5, we present arrival time prediction by the SAP model for 19 Earth-directed CME-ICME pairs and discuss its prediction ability. In section 6, we discuss the geomagnetic impact of super-massive, super-fast CMEs associated with solar superflares based on the SAP model.

\section{THE SHEATH-ACCUMULATING PROPAGATION (SAP) MODEL}
\subsection{BASIC ASSUMPTIONS}
In the SAP model, we assume the background solar wind as a spherically symmetric flow with uniform speed $V_{sw}$ in heliocentric location $r>r_0=0.1$AU. We express the total mass of the composite of the ICME and interplanetary sheath at time $t$ as $M(t)$, and the radial speed and the heliocentric distance of the center of mass as $V(t)$ and $R(t)$, respectively. Initially (at $t=0$), the total mass, radial speed and the heliocentric distance of the ICME-sheath composite are $M(0)=M_0$, $V(0)=V_0$ and $R(0)=r_0$, respectively (Figure 1). For simplicity, we call $V(t)$ and $R(t)$ the ICME speed and location, respectively, throughout this paper. We assume the ICME angular half width $\theta_0$ is constant during its propagation. In reality, the CME properties near the Sun are basically estimated by coronagraph observation. In the SAP model, we neglect the effect of gravitational and Lorentz force on the propagation of ICME in $r>r_0$ space.

The total mass of ICME-sheath composite at time $t$ is expressed as
\begin{equation}
M(t)=M_0+M_{sheath}(t)
\end{equation}
where $M_{sheath}(t)$ is the mass of interplanetary sheath ahead of ICME (Figure 1).
Assuming a constant fraction $c_0\simeq1$ of plasma swept by interplanetary shock becomes a part of interplanetary sheath, we obtain
\begin{equation}
M_{sheath}(t)=c_0\Omega_0\int_0^t\rho_{sw}(R(t'))R(t')^2 (V_{shock}(R(t'))-V_{sw})dt'
\end{equation}
where $\rho_{sw}(R)$ and $V_{shock}(R)$ are solar wind mass density and interplanetary shock propagation speed at $r=R$, respectively. The ICME solid angle $\Omega_0$ is approximated by the half angular width $\theta_0$ as $\Omega_0\simeq\pi\theta_0^2$, assuming a circular cross section of the ICME. In the SAP model, we think of a spherically symmetric wind i.e. $\rho_{sw}(R)R^2V_{sw}=\dot{M}_{sw}/4\pi$, where $\dot{M}_{sw}$ is a solar mass loss rate by the solar wind which is a constant.
Approximating the shock propagation speed by ICME speed, i.e. $V_{shock}\simeq{}dR/dt$, we get
\begin{equation}
M_{sheath}(t)=c_0\dot{M}_{sw}\frac{\Omega_0}{4\pi}\frac{R(t)-r_0-V_{sw}t}{V_{sw}}.
\end{equation}
On the other hand, the conservation of momentum of ICME-sheath composite is written as
\begin{equation}
(M_0+M_{sheath}(t))V(t)\simeq{}M_0V_0+M_{sheath}(t)V_{sw}.
\end{equation}

\subsection{THE ICME PROPAGATION CHARACTERISTICS IN THE SAP MODEL}
By solving Equation (3) and (4), we can express the ICME arrival time (t), the sheath mass ($M_{sheath}$), ICME speed ($V$) and the deceleration ($-a$) as a function of ICME heliocentric location $R$ as follows,
\begin{equation}
t(R)=\frac{R-r_0}{V_{sw}}\bigl\{1-\frac{M_0V_0}{M_c(R)V_{sw}}\epsilon(R)\bigr\},
\end{equation}
\begin{equation}
M_{sheath}(R)=M_0\frac{V_0}{V_{sw}}\epsilon(R),
\end{equation}
\begin{equation}
V(R)=(V_0-V_{sw})\bigg(1+\frac{V_0}{V_{sw}}\epsilon(R)\bigg)^{-1}+V_{sw},
\end{equation}
\begin{equation}
-a(R)=\frac{c_0\Omega_0\dot{M}_{sw}}{4\pi{}(M_0+M_{sheath}(R))V_{sw}}(V(R)-V_{sw})^2,
\end{equation}
where $M_c(R)$ and $\epsilon(R)$ are,
\begin{equation}
M_c(R)=c_0\Omega_0\int_{r_0}^{R}\rho_{sw}(R')R'^2dR'=c_0\dot{M}_{sw}\frac{\Omega_0(R-r_0)}{4\pi{}V_{sw}}
\end{equation}
\begin{equation}
\epsilon(R)=\sqrt{1+2\frac{M_c(R)V_{sw}(V_0-V_{sw})}{M_0V_0^2}}-1.
\end{equation}
The detailed derivation of the Equations (5)-(8) is given in the appendix A.

Generally, massive CMEs experience only a weak deceleration and although the the deceleration $-a(R)$ decays slower with $R$, they do not lose their speed so fast as light CMEs where $-a(R)$ is initially very strong (see Figure 1 of \citet{vrsnak2013} as an example in the drag-based model). Figure 2 (a)-(d) show $V(R)$, -$a(R)$, $t(R)$ and $M_{sheath}(R)/M_0$ with six different CME parameters in the SAP model.  When the CME mass is $M_0=3\times10^{15}$~g (thin lines), the ICME experience rapid deceleration before $R\sim0.3$AU followed by gradual deceleration afterwards. In that case, the ICME arrive at 1 AU with almost the same speed as background solar wind. This behavior is consistent with the two-phased deceleration characteristics of ICME reported by \citet{liu2013}.
When the CME is as heavy as $M_0=3\times10^{16}$~g (thick lines), the ICME arrives at 1 AU with larger speed compared with $M_0=3\times{}10^{15}$~g cases. We also note that the SAP model, as well as the drag-based model, expects that faster CMEs experience stronger deceleration. This tendency is consistent with observed deceleration of the shock front from the Sun to 1AU reported by \citet{woo1985}.
In section 2.3, we discuss the fundamental condition for fast ICMEs at Earth (e.g. $V({\rm{1~AU}})\gtrsim1000$km s$^{-1}$) as in the case of 2012 July 23 super-intense ESP event.

\subsection{THE FUNDAMENTAL CONDITION FOR EXTREMELY FAST ICME AT 1 AU}
First, we consider the heliocentric distance $R_c$ at which the relative speed of ICME with respect to the solar wind is halved i.e. $V(R_c)-V_{sw}=(V_0-V_{sw})/2$. From Equation (7), $R_c$ satisfies the following relation.
\begin{equation}
\epsilon(R_c)=\frac{V_0}{V_{sw}}
\end{equation}
From Equations (6) and (11), this leads,
\begin{equation}
M_{sheath}(R_c)=M_0
\end{equation}
From Equations (8), (12) and $V(R_c)-V_{sw}=(V_0-V_{sw})/2$, we get $a(R_c)=a(r_0)/8$, which means the rapid ICME deceleration almost ceases at $r=R_c$. We  call $R_c$ a ``{\it{deceleration cessation distance}}'', which is originally discussed in the empirical models \citep{gopalswamy2001}. 
Solving Equation (11), $R_c$ can be written in terms of $M_0$ and $V_0$ as follows,
\begin{equation}
R_c=r_0+({\rm{1~AU}}-r_0)\frac{M_0}{M_c({\rm{1~AU}})}(1+\frac{3}{2}\frac{V_{sw}}{V_0-V_{sw}})
\end{equation}
When the CME mass $M_0$ is larger than $M_c(\rm{1~AU})$, $R_c$ is always larger than $\rm{1~AU}$. This means CMEs heavier than $M_c(\rm{1~AU})$ with any initial speed will arrive at 1 AU without significant deceleration. In this sense, we call $M_c(\rm{1~AU})$ the ``{\it{critical CME mass}}'' for 1 AU travel.

Thin and thick lines in the Figure 3 (a) shows $R_c$ against $V_0$ with the CME mass of $M_0=3\times10^{15}$~g and $M_0=3\times{}10^{16}$~g, respectively in slow background solar wind. Figure 3 (b) shows $R_c$ with the same CME parameters as in panel (a) but in a fast background solar wind. $M_c({\rm{1~AU}})$ in the case of slow and fast winds are $M_c=9.1\times{}10^{16}$~g and $M_c=3.0\times{}10^{16}$~g, respectively. When the CME is heavier than or comparable to the critical mass (as in the case of the thick line in Figure 3 (b)), ICME will stay fast when arriving at Earth.

We compare the extremely fast ICME in 2012 July 23 event and the SAP model prediction.
Before the arrival of the ICME, solar wind speed and density measured in-situ by $STEREO-A$ at the distance of 0.96 AU from the Sun were roughly $V_{sw}({\rm{0.96 AU}})\simeq$500 km s$^{-1}$ and $n_{sw}({\rm{0.96 AU}})\simeq1-3$ cm$^{-3}$ with little variation. The corresponding critical CME mass is $M_c({\rm{0.96 AU}})=1\times10^{16}-3\times10^{16}$~g. The peak CME speed near the Sun is $V_0\simeq{}3050$ km s$^{-1}$ \citep{liu2014} and the CME mass estimated with $SOHO$/LASCO was about $M_0=3.2\times10^{16}$~g.
With these values, Equation (7) predicts the ICME speed at 0.96 AU to be between $2.0\times10^{3}$ km s$^{-1}$ and $2.6\times10^{3}$ km s$^{-1}$ which is consistent with the post-shock peak speed of 2250 km s$^{-1}$ measured at $STEREO-A$. We assumed $c_0=1$ and $\theta_0=\pi/4$ in above estimation.

\section{THICKNESS OF INTERPLANETARY SHEATH}
Although the thickness of the sheath is an important parameter for understanding space weather storms such as geomagnetic storms, they are rarely considered in modeling or observations. In this section, we discuss the thickness of the interplanetary sheath $D(R)$ expected by the SAP model. The fast mode Mach number of the leading shock is approximated as $\mathcal{M}_f(R)\simeq{}(V(R)-V_{sw})/C_f(R)$, with $C_f$ being the phase speed of fast mode MHD wave in background solar wind plasma. On the other hand, the sheath mass is approximately written as,
\begin{equation}
M_{sheath}(R)\simeq\chi(R)\rho_{sw}(R)\Omega_0{}R^2D(R)=\frac{\Omega_0}{4\pi}\frac{\dot{M}_{sw}\chi(R)D(R)}{V_{sw}}
\end{equation}
where $\chi(R)$ is the compression ratio of the interplanetary shock which depends not only on $\mathcal{M}_f(R)$ but also on upstream plasma beta and the angle between the shock normal and upstream magnetic field. 
From Equations (6) and (14), $D(R)$ can be approximated as follows,
\begin{equation}
D(R)\simeq\frac{4\pi{}M_0V_0}{\Omega_0{}\dot{M}_{sw}}\frac{\epsilon(R)}{\chi(R)}.
\end{equation}

If we assume the value of $M_0$, $V_0$, $V_{sw}({\rm{0.96 AU}})$ and $n_{sw}({\rm{0.96 AU}})$ as those assumed in the previous section for 2012 July 23 event, and also $\chi\simeq3$, the sheath thickness estimated with Equation (15) become $D(0.96 AU)\simeq0.25$ AU. The estimated sheath thickness of $\sim0.25$ AU is substantially larger than the typical thickness of $\sim0.05$ AU reported by \citet{russell2002} because of exceptionally large $V_0$ and $M_0$ in 2012 July 23 event. Assuming the sheath speed during its passage at the spacecraft to be $\simeq 2\times10^3$ km s$^{-1}$, the transit time of the sheath of thickness 0.25 AU would be $\simeq$ 5 hours. The actual solar wind disturbance detected by $STEREO-A$ is known to be the merger of two successively launched CMEs \citep{liu2014}. The leading edges of the two CMEs passed the spacecraft after the arrival of the leading shock front by 2 and 6 hours, respectively. The predicted sheath transit time of 5 hours is in between the two but more consistent with the latter.

\section{COMPARISON OF THE SAP MODEL WITH THE DRAG-BASED MODEL AND THE `SNOW-PLOUGH' MODEL OF \citet{tappin2006}}
From the relation $\dot{M}_{sw}=4\pi\rho_{sw}(R)R^2V_{sw}$, the ICME acceleration in the SAP model (Equation (8)) can be expressed as follows,
\begin{equation}
a=-\frac{c_0A\rho_{sw}}{M_0+M_{sheath}}(V-V_{sw})^2
\end{equation}
with $A=\Omega_0R^2$ being the cross sectional area of ICME.

On the other hand, the CME acceleration in the drag-based model is as follows,
\begin{equation}
a=-\frac{c_dA\rho_{sw}}{M_0+M_{\mathcal{V}}}(V-V_{sw})|V-V_{sw}|,
\end{equation}
where $c_d$ is a drag coefficient of order unity and $M_{\mathcal{V}}=\rho_{sw}\mathcal{V}/2$ is a ``virtual mass'' with $\mathcal{V}$ being instantaneous CME volume (\citet{vrsnak2013}; see also \citet{cargill2004} and references therein). 

The ``virtual mass'' formulation incorporated in the drag-based model is based on the assumption of potential flow passed a solid sphere \citep{landau1959}. However, when $V-V_{sw}$ is transonic or supersonic, the flow around the CME would be substantially different from the potential flow due to strong compressibility (e.g. sheath forms ahead of the CME). In such a case, the SAP model would give a straightforward estimation of the virtual mass ($M_{\mathcal{V}}=M_{sheath}$).

The dynamics of sheath accumulation discussed in the SAP model is basically close to the ``piston-driven'' shock formation process \citep{vrsnak2008}. On the basis of the piston-driven mechanism, the shock-driving CME is not necessarily supersonic. \citet{sheeley1985} reported that the shocks tend to be associated with faster CMEs (with their speeds larger than 500 km s$^{-1}$), while sometimes associated with slower CMEs with speeds between 200-400 km s$^{-1}$. When the sheath thickness is comparable to or larger than the lateral extent of the CME i.e. $D(R)\gtrsim R\theta_0$, a large part of the shocked plasma would escape from the sides of the CME, deviating from the piston-driven mechanism. In that case, the drag-based model rather than the SAP model would give more appropriate description of the CME deceleration. The typical thickness of the sheath at 1 AU reported by \citet{russell2002} is $\sim 0.05$ AU, which is likely much smaller than typical widths of CMEs at 1AU of $\sim 1AU$ assuming $\theta_0\sim1$. From this, we expect the SAP model can be widely applied to the decelerating propagation of CMEs.

The ``snow-plough'' model proposed by \citet{tappin2006} is mathematically very similar to the SAP model. The detailed comparison between the formulas of the two are given in Appendix B. The SAP model is basically different to the ``snow-plough'' model in that it tracks the evolution of the sheath and that it is an analytical model with the assumption of the uniform solar wind speed.

\section{1 AU ARRIVAL TIME PREDICTION BASED ON THE SAP MODEL}
\citet{makela2016} studied the relation between radial speed and expansion speed of 19 Earth directed CMEs that occurred during January 2010 to September 2012 when the $STEREO$ and $SOHO$ were viewing the Sun in near quadrature. We apply the SAP model to predict 1 AU arrival time of the same set of CME-ICME pairs as studied in \citet{makela2016}. The average angular half width of the 19 CMEs measured by STEREO/COR2 is $\bar{\theta_0}=0.23\pi$. The mass is estimated for 16 out of 19 CMEs and listed in online SOHO/LASCO CME catalog \footnote{http://cdaw.gsfc.nasa.gov/CME\_list/} \citep{yashiro2004}. The average CME mass of the 16 CMEs is $\bar{M_0}=9.6\times{}10^{15}$~g. We note that the mass estimation of the Earth-directed CMEs based on $SOHO$/LASCO data is based on many assumptions which possibly introduce significant uncertainty in mass estimation. We refer to the observed 1 AU arrival time of interplanetary shock as $t_{obs}$. We refer to the predicted 1 AU arrival time based on the SAP model as $t_{SAP}$. 
\begin{equation}
t_{SAP}=t_{in}+t({\rm{1~AU}})-\frac{1}{2}t_{s}
\end{equation}
where $t_{in}\simeq{}r_0/V_0$ is time for a CME to travel from the Sun to $r=r_0$, and $t_s=D({\rm{1~AU}})/V({\rm{1~AU}})$ is  sheath transit time at Earth. We assumed $M_{sheath}({\rm{1~AU}}) >> M_0$ so the ICME center of mass when $R(t)=\rm{1~AU}$ is almost at the midpoint of the sheath region. The average of the mass ratio $M_{sheath}/M_0$ expected from the SAP model using Equation (6) is 5.5. We assumed typical slow background solar wind with $V_{sw}=350$ km s$^{-1}$ and $n_{sw}({\rm{1~AU}})=9$ cm$^{-3}$ \citep{schwenn2006}. $M_0=\bar{M_0}$ and $\theta_0=\bar{\theta_0}$ are assumed in the calculation of $t({\rm{1~AU}})$, so $t_{SAP}$ only depends on initial CME speed $V_0$ in this study. In the evaluation of $t_s$, we assumed shock compression ratio to be $3$ in all cases for simplicity. We tried three different values of $c_0$ of 0.8, 0.9 and 1.0, and found $c_0=0.9$ minimized the root-mean-square ($RMS$) of the observed-minus-calculated transit time differences. We note that we cannot draw the conclusion merely from this that 90 percent of shocked solar wind plasma have been actually accumulated in the sheath in average, partly because assumed CME parameters, especially the mass, contains large uncertainty.The $RMS$ and the maximum value of the observed-minus-calculated transit time differences in $c_0=0.9$ case were 7.5 hours and 15.8 hours, respectively. Figure 3 shows the correlation plot between predicted ($t_{SAP}$) and observed ($t_{obs}$) arrival times in $c_0=0.9$ case.

\section{SUPER-INTENSE GEOMAGNETIC STORMS ASSOCIATED WITH SOLAR SUPERFLARES}
The ICME-driven westward electric field at Earth ($E_y=VB_s$, with $V$ and $B_s$ being the speed and southward magnetic field of ICME) is the crucial space plasma quantity that drive intense geomagnetic storms \citep{burton1975,yermolaev2007}. Magnetic cloud core field is known to be correlated with the ICME speed \citep{gonzalez1998} and the upper limit can be estimated by the equipartition field strength as follows \citep{takahashi2016}, 
\begin{equation}
B_{s,upperlimit}({\rm{1~AU}})\simeq\sqrt{4\pi\rho_{sw}({\rm{1~AU}})}(V({\rm{1~AU}})-V_{sw}).
\end{equation}
The upper limit of $E_y$ at 1 AU can be expressed by $B_{s,upperlimit}(\rm{1~AU})$ and $V(\rm{1~AU})$ as follows,
\begin{equation}
E_{y,upperlimit}({\rm{1~AU}})\simeq{}V({\rm{1~AU}})B_{s,upperlimit}({\rm{1~AU}})
\end{equation}
From equation (19) and (20), $E_{y,upperlimit}({\rm{1~AU}})$ is determined solely by ICME speed (and background solar wind density) at 1 AU, which depends both on the mass and speed of causative CMEs due to Equation (7).

Based on the scaling relations among CME properties and flare soft X-ray peak flux ($F_{SXR}$) at 0.1nm-0.8nm measured by X-ray detector on board $GOES$ satellite discussed in \citet{takahashi2016}, on the other hand, the upper limit of the mass and speed of CMEs are expressed as follows,
\begin{equation}
M_{0,upperlimit}\simeq3\times10^{16}\times\bigg(\frac{F_{SXR}}{F_{SXR,X10}}\bigg)^{2/3}~\rm{g}
\end{equation}
\begin{equation}
V_{0,upperlimit}\simeq4.2\times10^3\times\bigg(\frac{F_{SXR}}{F_{SXR,X10}}\bigg)^{1/6}\rm{km~s^{-1}}
\end{equation}
with $F_{SXR,X10}=0.001$ W m$^{-2}$ being $F_{SXR}$ of X10 class flare.
In Equation (21), we assumed the maximum CME mass associated with X10 flare to be $\sim3\times{}10^{16}$~g \citep{aarnio2012}.
Applying Equations (21) and (22) for the evaluation of $E_{y,upperlimit}({\rm{1~AU}})$ in Equation (20), we get $E_{y,upperlimit}({\rm{1~AU}})$ as a function of $F_{SXR}$. 
$E_{y,upperlimit}({\rm{1~AU}})$ against $F_{SXR}$ (or flare class) with fast and slow background solar wind are plotted in Figure 5. When $M_{0,upperlimit}\gtrsim{}M_c({\rm{1~AU}})$, $V({\rm{1~AU}})-V_{sw}\simeq{}V_0$ holds, and $E_{y,upperlimit}(\rm{1~AU})$ approximately scales as $E_{y,upperlimit}({\rm{1~AU}})\propto{}V_0^2\propto{}F_{SXR}^{1/3}$. Based on the discussion above, $E_{y,upperlimit}$ associated with X10 flare, for example, will be $\sim2\times10^3$~mV~m$^{-1}$. If such $E_y$ continues for $\sim{}2$ hours, $Dst$ would be $Dst\sim-2\times10^3$~nT, following the formula by \citet{burton1975}. On the other hand, the upper limit of $-Dst$ inherent in geomagnetism is evaluated to be $\sim2500$ nT in \citet{vasyliunas2011}, which is comparable to the upper limit of $-Dst$ associated with X10 flare above. Further careful discussion is needed to evaluate actual upper limit of $-Dst$ associated with huge solar flares of $\gtrsim$X10 class.

\appendix
\section{THE DERIVATION OF EQUATIONS (5)-(8)}
In the appendix, we derive Equations (5)-(8) from Equations (3) and (4).

We first define two new variables, $\tilde{R}(t)=R(t)-r_0-V_{sw}t$ and $\tilde{V}(t)=d\tilde{R}(t)/dt=V(t)-V_{sw}$, so that Equations (3) and (4) can be expressed as
\begin{equation}
M_{sheath}(t)=c_0\dot{M}_{sw}\frac{\Omega_0}{4\pi}\frac{\tilde{R}(t)}{V_{sw}}
\end{equation}
\begin{equation}
(M_0+M_{sheath}(t))(\tilde{V}(t)+V_{sw})=M_0V_0+M_{sheath}(t)V_{sw}
\end{equation}
Solving equation (A2) with respect to $\tilde{V}(t)$, we get,
\begin{equation}
\tilde{V}(t)=\frac{M_0}{M_0+M_{sheath}(t)}\tilde{V}_0,
\end{equation}
with $\tilde{V}_0=V_0-V_{sw}$.
Substituting Equation (A1) into Equation (A3), we get,
\begin{equation}
\tilde{V}=\frac{d\tilde{R}(t)}{dt}=\bigg(1+\frac{c_0\Omega_0\dot{M}_{sw}\tilde{R}(t)}{4\pi{}M_0V_{sw}}\bigg)^{-1}\tilde{V}_0
\end{equation}
Integrating Equation (A4) by $t$, we get
\begin{equation}
\tilde{R}(t)+\frac{c_0\Omega_0{}\dot{M}_{sw}}{8\pi{}M_0V_{sw}}\tilde{R}(t)^2=\tilde{V}_{0}t
\end{equation}
Substituting $\tilde{R}=R-r_0-V_{sw}t$ into Equation (A5), we get a quadratic equation of $t$ as follows,
\begin{equation}
(R-r_0-V_{sw}t)+\frac{c_0\Omega_0{}\dot{M}_{sw}}{8\pi{}M_0V_{sw}}(R-r_0-V_{sw}t)^2=(V_0-V_{sw})t
\end{equation}
Solving the quadratic Equation (A6) with respect to $t$, we get the arrival time t(R) as given by Equation (5), where we used abbreviations $\epsilon$ and $M_c$ defined by Equations (9) and (10).

Then substituting Equation (5) into $\tilde{R}=R-r_0-V_{sw}t$, we express $\tilde{R}$ in terms of $R$ as follows,
\begin{equation}
\tilde{R}=(R-r_0)\frac{M_0V_0\epsilon(R)}{M_c(R)V_{sw}}
\end{equation}
If we substitute Equation (A7) into Equation (A1), we get the sheath mass $M_{sheath}(R)$ as Equation (6). 

Substituting Equation (A7) into Equation (A4), we get $V=\tilde{V}+V_{sw}$ as in Equation (7).

Lastly, we derive Equation (8).
The acceleration $a$ can be deformed as follows,
\begin{equation}
a=\frac{dV}{dt}=\frac{d\tilde{V}}{dt}=\frac{d\tilde{V}}{d\tilde{R}}\frac{d\tilde{R}}{dt}=\tilde{V}\frac{d\tilde{V}}{d\tilde{R}}
\end{equation}
On the other hand, making the derivative of Equation (A4) with respect to $\tilde{R}$, we get,
\begin{equation}
\frac{d\tilde{V}}{d\tilde{R}}=-\frac{c_0\Omega_0\dot{M}_{sw}\tilde{V_0}}{4\pi{}M_0V_{sw}}\bigg(1+\frac{c_0\Omega_0\dot{M}_{sw}\tilde{R}}{4\pi{}M_0V_{sw}}\bigg)^{-2}
\end{equation}
Substituting Equations (A1) and (A4) into Equation (A9), we get,
\begin{equation}
\frac{d\tilde{V}}{d\tilde{R}}=-\frac{c_0\Omega_0\dot{M}_{sw}}{4\pi{}(M_0+M_{sheath})V_{sw}}\tilde{V}
\end{equation}
Substituting Equation (A10) into Equation (A8), we get ICME deceleration $-a$ as Equation (8).

\section{THE COMPARISON BETWEEN THE `SNOW-PLOUGH' MODEL OF Tappin (2006)  AND THE SAP MODEL}
The `snow-plough' model proposed in \citet{tappin2006} is a set of 2 coupled differential equations:
\begin{equation}
\frac{dV_t}{dt}=-\frac{dM_t}{dt}\frac{V_t-V_{sw}}{M_t}
\end{equation}
\begin{equation}
\frac{dM_t}{dt}=\Omega_0 \rho_{sw} R_t^2 (V_t-V_{sw})
\end{equation}
where $R_t$, $M_t$ and $V_t$ are the heliocentric distance, the mass and the speed of a transient. The transient gets heavier by sweeping up solar wind plasma ahead of it.

On the other hand, the SAP model is based on the combination of Equations (3) and (4). 
Making the time derivative of Equation (3) and using the relations $\rho_{sw}R^2 V_{sw}=\dot{M}_{sw}/{4\pi}$, and $M=M_0+M_{sheath}$, we get the following differential equation: 
\begin{equation}
\frac{dM}{dt}=c_0\Omega_0 \rho_{sw} R^2 (V-V_{sw})
\end{equation}
Then, if we make the time derivative of Equation (4) with slight deformation , we get
\begin{equation}
\frac{dV}{dt}=-\frac{dM}{dt}\frac{V-V_{sw}}{M}.
\end{equation}
If we assume $M=M_t$ and $c_0=1$ in Equation (B3) of the SAP model, we get Equation (B2) of the `snow-plough' model.
If we further assume $V=V_t$, Equation (B4) become equivalent to Equation (B1).

The authors are grateful to the journal referees for their comments that have made the paper much clearer. This work is financially supported by the Grant-in-Aid for JSPS Fellows 15J02548 and JSPS KAKENHI Grant Numbers 16H03955. Also, T. T. is financially supported from the Unit of Synergetic Studies for Space in Kyoto University.



\begin{figure}
\epsscale{.90}
\plotone{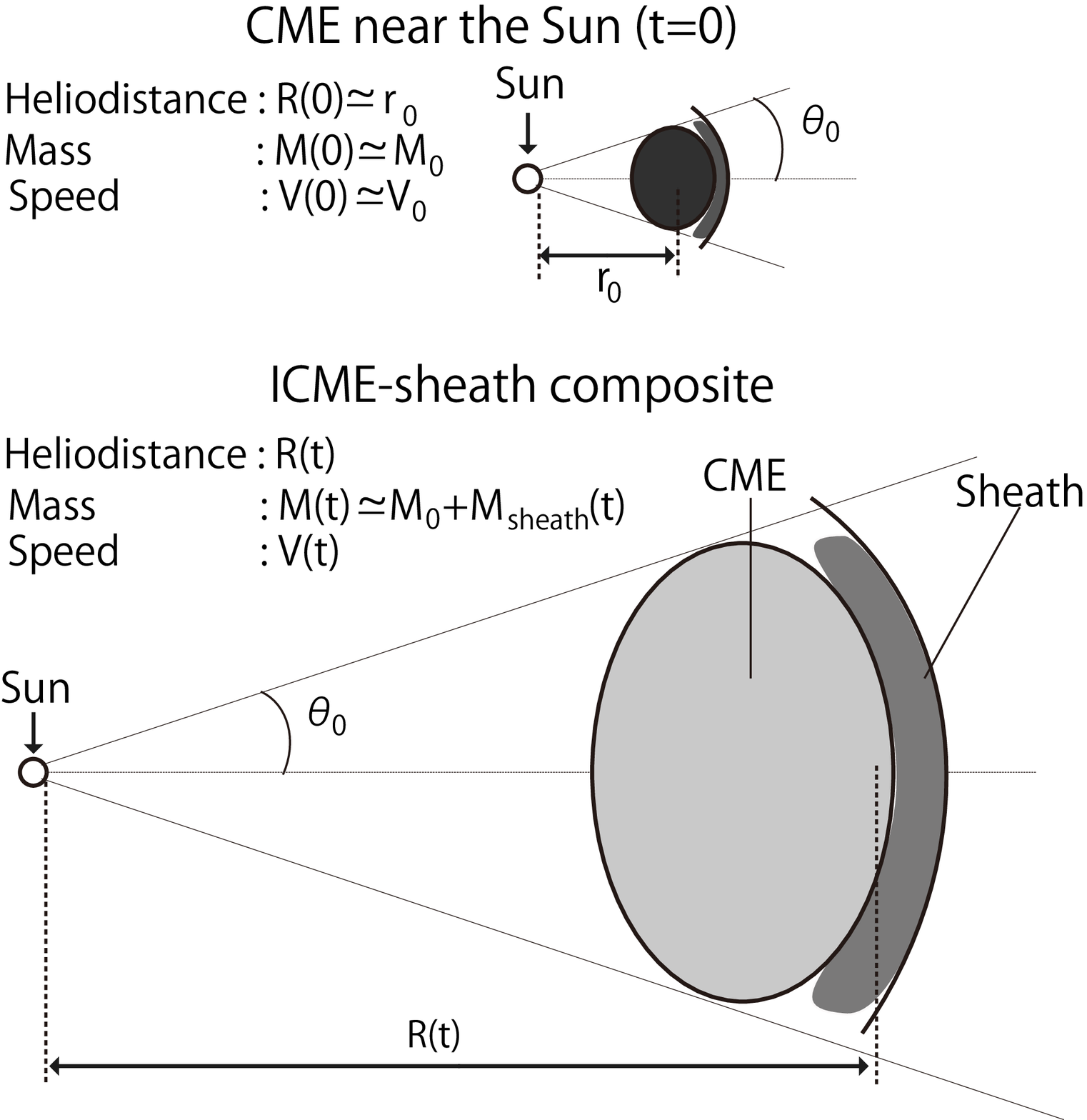}
\caption{The schematic figure of the SAP model. In the SAP model, the initial CME parameters ($M_0$, $V_0$ and $\theta_0$) are given when the center of mass of CME is at the heliocentric location $r=r_0$. During the propagation in the interplanetary space, the total mass of ICME-sheath composite is approximated by the sum of the CME mass and the sheath mass ($M(t)\simeq{}M_{0}+M_{sheath}(t)$). The heliocentric location and radial speed of the ICME at time $t$ are $R(t)$ and $V(t)$, respectively.
\label{flare}}
\end{figure}

\begin{figure}
\epsscale{.90}
\plotone{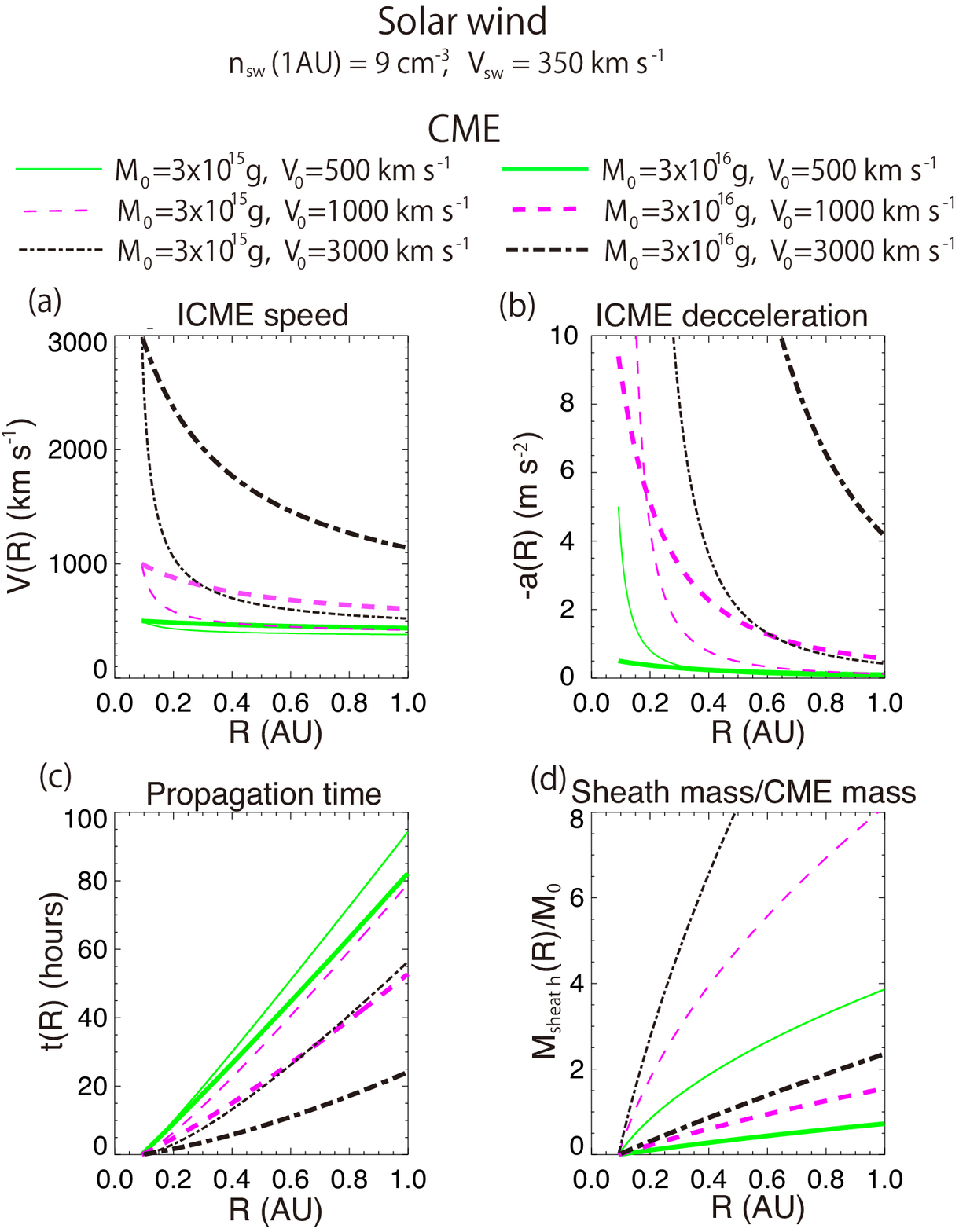}
\caption{ICME propagation properties in the case of slow background solar wind as a function of ICME heliocentric distance $R$. ICME speed ($V(R)$), deceleration ($-a(R)$), arrival time ($t(R)$) and sheath mass in units of initial CME mass $M_{sheath}(R)/M_0$ are plotted in panels (a) to (d), respectively with six different pairs of $M_0$ and $V_0$ values. The CME mass is chosen to be $M_0=3\times10^{15}$ g or $3\times10^{16}$ g, while the CME speed is takes three values, that are $V_0=500$ km s$^{-1}$, 1000 km s$^{-1}$ or $3000$ km s$^{-1}$. $\theta_0=\pi/4$ and $c_0=1$ are assumed in all the cases.
\label{flare}}
\end{figure}

\begin{figure}
\epsscale{.90}
\plotone{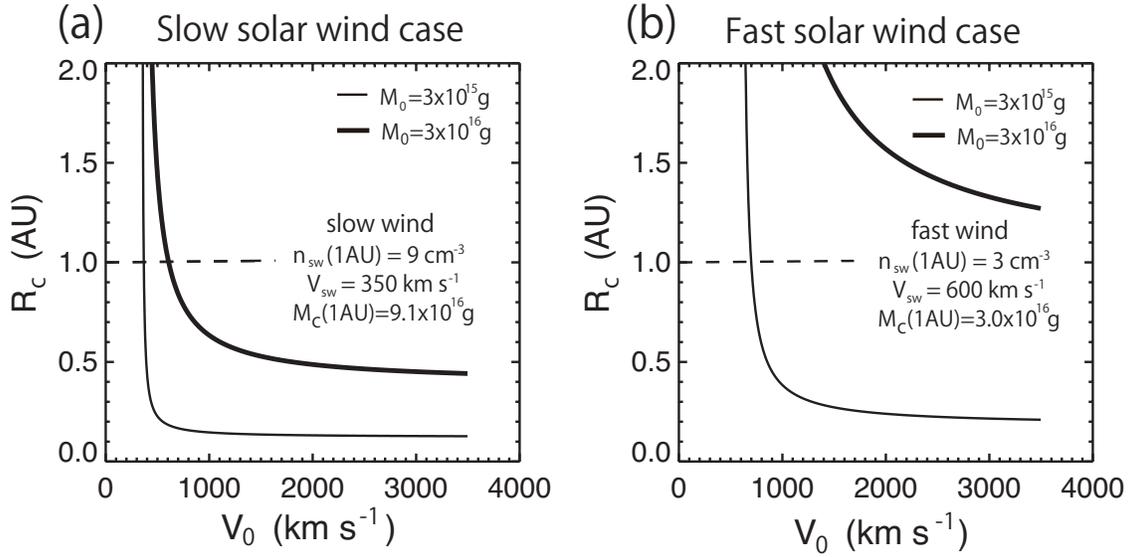}
\caption{$R_c$ against $V_0$ in slow (panel (a))and fast (panel (b)) background solar wind. Thin and solid lines show $R_c$ in the cases of heavy ($M_0=3\times10^{15}$~g) and super heavy ($M_0=3\times10^{16}$~g) CMEs, respectively. The solar wind density and speed at 1AU are chosen to be $n_{sw}(1AU)=9$ cm$^{-3}$ and $V_{sw}(1AU)=350$ km s$^{-1}$ for slow wind case, and $n_{sw}(1AU)=3$ cm$^{-3}$ and $V_{sw}(1AU)=600$ km s$^{-1}$ for fast wind case, respectively. $M_c(\rm{1~AU})$ for slow and fast wind cases are $M_c=9.1\times10^{16}$~g and $3.0\times10^{16}$~g, respectively. $\theta_0=\pi/4$ and $c_0=1$ are assumed in all cases.
\label{vp}}
\end{figure}


\begin{figure}
\epsscale{.90}
\plotone{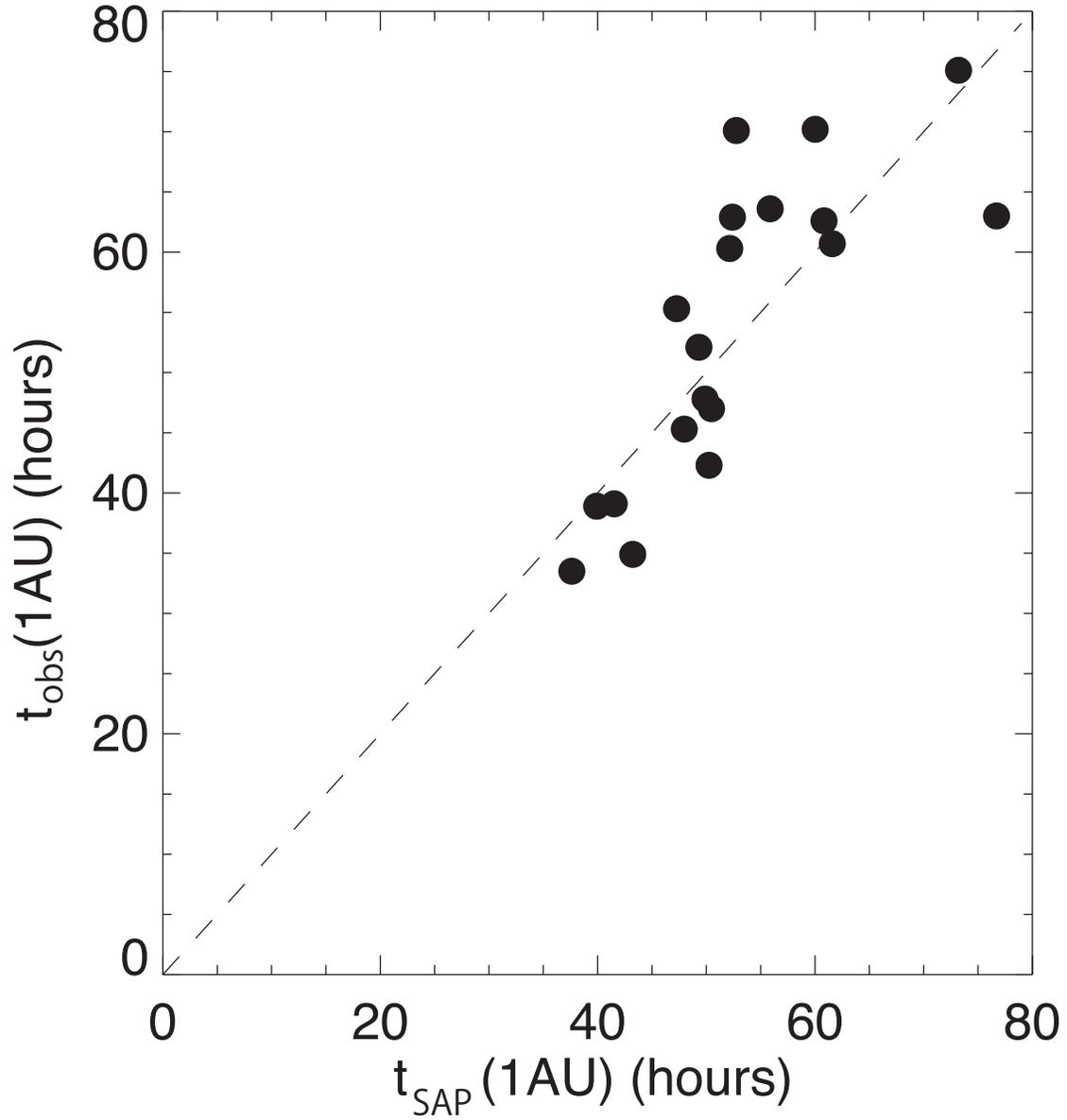}
\caption{Predicted v.s. observed 1AU arrival time. The relation between observed 1 AU arrival times ($t_{obs}$(1 AU)) and those predicted by the SAP model ($t_{SAP}$(1 AU)) for the 19 CME-ICME pairs are plotted as a correlation plot.
\label{flare}}
\end{figure}

\begin{figure}
\epsscale{.90}
\plotone{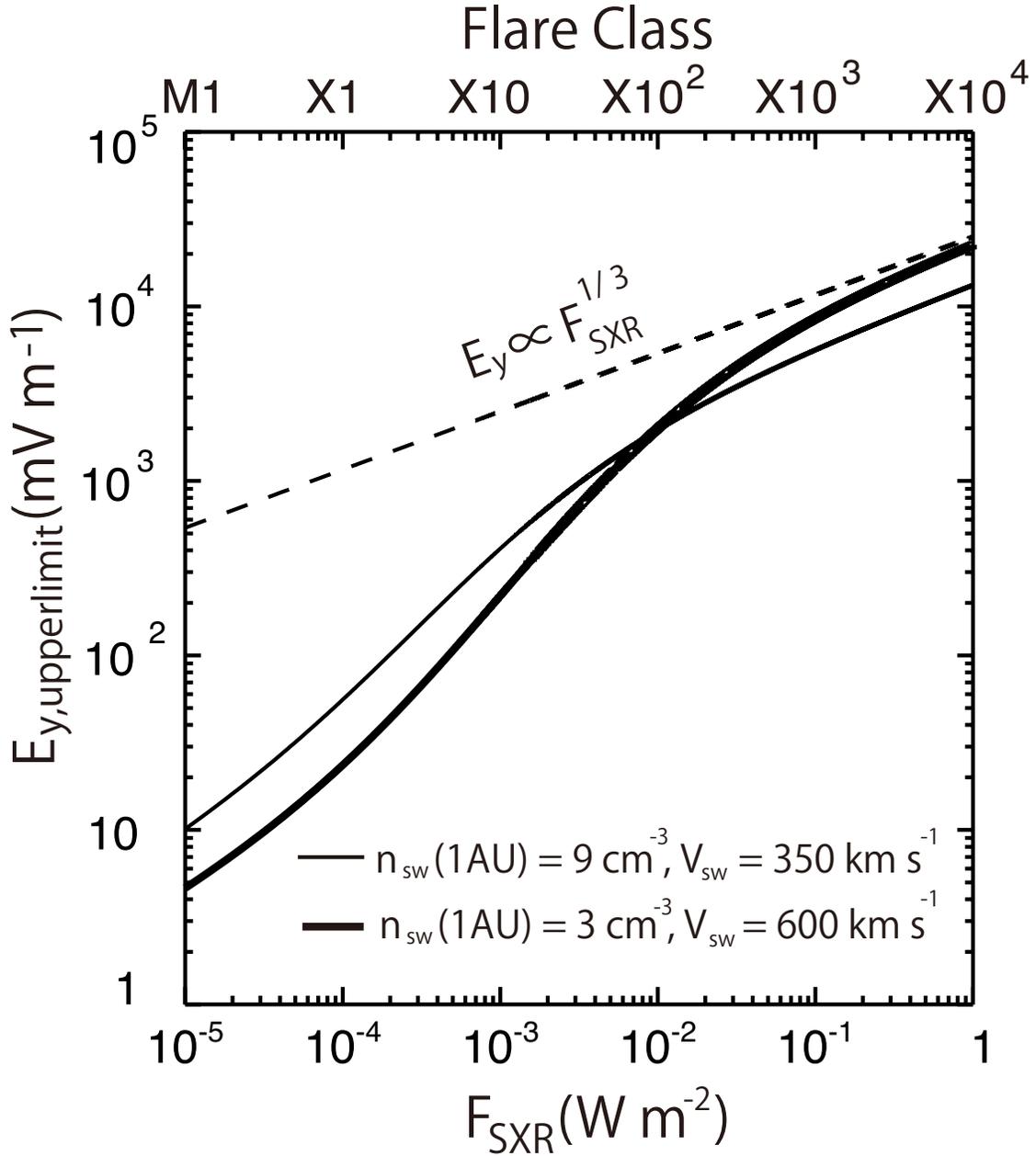}
\caption{The upper limit of westward electric field ($E_{y,upperlimit}$) against the SXR peak flux of associated flare ($F_{SXR}$). Thin and thick lines denote the cases with slow and fast background solar wind, respectively. The asymptotic line with the slope of $E_{y}\propto{}F_{SXR}^{1/3}$ is plotted as a dashed line.
\label{flare}}
\end{figure}

\end{document}